\documentclass[aps,preprint,preprintnumbers,showpacs,nofootinbib]{revtex4}
\usepackage{graphicx}

\begin{document}
\title{Zee model and Neutrinoless double beta decay}
\author{Ming-Yang Cheng$^1$ and Kingman Cheung$^2$}
\email[Email:]{cheung@phys.cts.nthu.edu.tw}
\affiliation{
$^1$ Department of Physics, National Tsing Hua University, Hsinchu, Taiwan, 
R.O.C. \\
$^2$ National Center for Theoretical Sciences, Hsinchu, Taiwan, R.O.C.}
\date{\today}

\begin{abstract}
The original Zee model can easily accommodate a bi-maximal mixing
 solution for the atmospheric and solar neutrino problems.  From 
the most recent fit to the data we obtain a set of 
parameters for the Zee model.  However, the recent claim on a positive signal
observed in a neutrinoless double beta decay, which requires 
a nonzero $m_{ee}$, cannot be accounted for by the Zee model.  
Based on the observed neutrino mass-square differences and mixings, 
we derive a few general 
patterns for the neutrino mass matrix in the flavor basis, which can correctly
describe the atmospheric, solar, and $0\nu\beta\beta$ data. Finally, we 
investigate a few possible extensions to the Zee model, which can
give the required mass patterns.
\end{abstract}
\preprint{NSC-NCTS-020304}
\maketitle

\section{Introduction}

More and more neutrino data are accumulated since a few years ago.  They all
contribute to the understanding of neutrino masses and mixings.  They are
very important clues to the underlying theory for neutrino masses, and perhaps
to other fermions as well.  It is therefore very important to pin down the
necessary patterns for neutrino mass matrix that can explain all the 
observations.  

Deficits in both atmospheric and solar neutrinos have been known for a long
time.  A possible and favorable explanation is neutrino oscillation,
which is made possible because neutrinos have different masses
and the flavor basis is not the same as the mass basis.
The atmospheric neutrino data from SuperK \cite{superk} and the preliminary
data from K2K \cite{k2k} showed a maximal mixing between the $\nu_\mu$ and
$\nu_\tau$ \cite{atm-fits} with a mass-square difference and a mixing angle:
\begin{equation}
\label{atm}
\Delta m^2_{\rm atm} \approx 3.0 \times 10^{-3}\;\; {\rm eV}^2\,, \;\;\;
\sin^2 2\theta_{\rm atm} = 0.99 \;.
\end{equation}
On the other hand, all the combined solar neutrino data favor
the oscillation of the $\nu_e$ into a mixture of $\nu_\mu$ and $\nu_\tau$.
Especially, the recent
SNO \cite{sno} data provided a convincing evidence that there is a non-$\nu_e$ 
component in the neutrino flux coming from the sun.  The most recent fit
favors, at least statistically, the large mixing angle (LMA) solution, namely,
\cite{solar-fits}
\begin{equation}
\label{solar}
\Delta m^2_{\rm sol} \approx 4.5 \times 10^{-5}\;\; {\rm eV}^2\,; \;\;\;
\tan^2 \theta_{\rm sol} = 0.41 \;,
\end{equation}
which gives $\sin^2 2\theta_{\rm sol} \approx 0.8$. 
The above two mass-square
 differences can be accommodated by 3 active neutrinos. 
The much debated result from LSND \cite{lsnd}, if confirmed, requires an
additional neutrino, which is sterile.  

Another recent result from a neutrinoless double beta decay ($0\nu\beta\beta$) 
experiment \cite{0nbeta} added another
clue or constraint to the neutrino mass matrix.  The positive signal
in the $0\nu\beta\beta$ decay implies a nonzero
$m_{ee}$ entry in the neutrino mass matrix arranged in the flavor basis:
\begin{displaymath}
\left(  \begin{array}{ccc}
            m_{ee}    & m_{e\mu}  & m_{e\tau} \\
            m_{e\mu}  & m_{\mu\mu}& m_{\mu\tau}  \\
            m_{e\tau} & m_{\mu\tau} &m_{\tau\tau}  \\
       \end{array} \right ) \;.
\end{displaymath}
The 95\% allowed range of $m_{ee}$ is $0.05 - 0.86$ eV with a best value of
$0.4$ eV.  It also implied that the electron neutrino is majorana in nature.
There is, however, an argument against this claim: see Ref. \cite{vissani,aal}.

Such a large value for $m_{ee} \gg \sqrt{\Delta m^2_{\rm atm,sol}}$, if true, 
gives a nontrivial modification to the neutrino mass patterns and mixings.  
A lot of possible
mass textures that were proposed to explain the atmospheric and solar
neutrino deficits become incompatible with the new $0\nu\beta\beta$ data.
A well-known example is the Zee model \cite{zee} that can generate a bi-maximal
mixing between the $\nu_\mu$ and $\nu_\tau$ and between
the $\nu_e$ and the mixture of $\nu_\mu$ and $\nu_\tau$ \cite{jarl,paul1,otto}
\footnote{For a number of analyses on the Zee model after the SNO data but 
before the $0\nu\beta\beta$ data, please see Refs. 
\cite{paul2,koide1,haba,brah,dan1}},
however, the Zee model guarantees the diagonal mass matrix elements to be zero.
Therefore, it cannot explain the nonzero $m_{ee}$ implied by the
$0\nu\beta\beta$ signal.  It would then be necessary to consider additional
new physics to the Zee model in order to accommodate the $m_{ee}$ data.
In this work, we look into the general mass patterns that can 
accommodate all the data, and from the patterns we introduce additional new 
physics to the Zee model.

The organization is as follows.  In Sec. II, we describe the Zee model 
and how it could
explain the atmospheric and solar neutrino data.  Using the most recent
fit to the data we obtain a set of values for the parameters of the Zee model.
In Sec. III, we derive the general mass patterns that can explain all the 
data.  In Sec. IV, we point out a few simple extensions to the Zee model that
can generate the necessary patterns.  We conclude in Sec. V.

\section{Zee model}

An economical way to generate small neutrino masses with a 
phenomenologically favorable texture is given by the Zee model
\cite{jarl,paul1,otto}, which generates masses via one-loop diagrams.
The model consists of a charged gauge singlet scalar $h^-_{\rm Zee}$, 
the Zee scalar, which couples to lepton doublets 
$\psi_{L}$ via the interaction
\begin{equation} 
f_{ab} \left( \psi^i_{a L} {\cal C} \psi^j_{b L} \right ) \;
\epsilon_{ij}\; h^-_{\rm Zee} \; +h.c. \;,
\end{equation}
where $i,j$ are the SU(2) indexes, $a,b$ are the generation indexes, 
${\cal C}$ is the charge-conjugation matrix, 
and $f_{ab}$ are Yukawa couplings antisymmetric in $a$ and $b$.  
Another ingredient of the Zee model is an extra Higgs doublet (in addition to 
the one that gives masses to charged leptons) that develops a 
vacuum expectation value (VEV) and thus provides mass mixing between
the charged Higgs boson and the Zee scalar boson.  Let us parametrize this
mixing term as 
\begin{equation}
   m_3^2 \,  h_d^+ \; h^-_{\rm Zee} + h.c. \;,
\end{equation}
where $m_3$ is of dimension $[{\rm mass}]$.
This coupling, together with the $f_{ab}$'s, enforces lepton-number violation.
Through one-loop diagrams the Zee model can generate off-diagonal majorana 
mass terms, $m_{ab} (a,b=e,\mu,\tau)$,  given by \cite{zee,otto,petcov}
\begin{equation}
\label{mab}
m_{ab} = - \frac{1}{16 \pi^2}\, f_{ab} \, 
\frac{g m_3^2}{\sqrt{2} M_W \cos\beta}\, (m_b^2 - m_a^2) \, 
h( M^2_{h_d^-}, M^2_{h^-_{\rm Zee}} )\;,
\end{equation}
where the function $h(x,y) = \log(y/x)/(x-y)$ and the charged Higgs $h_d^-$
couples to the charged lepton $l$ 
with a coupling $g m_l/(\sqrt{2}M_W \cos\beta)$.
It is a generic feature of the Zee model that the diagonal terms $m_{aa}=0$.  
Thus, we obtain a neutrino mass matrix of the following form
\[
 \left( \begin{array}{ccc}
0        & m_{e\mu} & m_{e\tau} \\
m_{e\mu} & 0        & m_{\mu\tau}  \\
m_{e\tau}& m_{\mu\tau} & 0 
         \end{array}          \right ) \; .
\]

Recent analyses \cite{jarl,paul1,otto,paul2,koide1,haba,brah,dan1,john,koide2}
showed that the Zee mass matrix of the following texture 
\[
\label{zee-m}
 \left( \begin{array}{ccc}
0        & m_{e\mu} & m_{e\tau} \\
m_{e\mu} & 0        & \epsilon  \\
m_{e\tau}& \epsilon & 0 
         \end{array}          \right ) \; ,
\]
where $\epsilon$ is small compared with $m_{e\mu}$ and $m_{e\tau}$,
is able to provide a compatible mass pattern that explains the atmospheric
and solar neutrino data.  
Moreover, $m_{e\mu} \sim m_{e\tau}$ is required to
give the maximal mixing solution for the atmospheric neutrinos.

Keeping the leading power in $\epsilon$ we obtain the mass eigenvalues
$m_1,m_2,m_3$ and the diagonalization matrix $U$ such that 
$U^T M_\nu U = diag\{m_1,m_2,m_3\}$
(we work in the basis where the charged-lepton mass matrix is diagonal):
\begin{eqnarray}
m_1 &=& \sqrt{ m^2_{e\mu} + m^2_{e\tau} } + \epsilon\, \frac{m_{e\mu} m_{e\tau}
}{m^2_{e\mu} + m^2_{e\tau}} \;,\nonumber \\
m_2 &=& - 
\sqrt{ m^2_{e\mu} + m^2_{e\tau} } + \epsilon\, \frac{m_{e\mu} m_{e\tau}
}{m^2_{e\mu} + m^2_{e\tau}} \;,\nonumber \\
m_3 &=& -2\epsilon \, \frac{m_{e\mu} m_{e\tau}
}{m^2_{e\mu} + m^2_{e\tau}} \;,\nonumber \\
U &=& \left( 
  \begin{array}{ccc}
\frac{1}{\sqrt{2}} \left( 1- \epsilon \frac{m_{e\mu} m_{e\tau} }{2 m^3}\right)&
\frac{1}{\sqrt{2}} \left( 1+ \epsilon \frac{m_{e\mu} m_{e\tau} }{2 m^3}\right)&
\epsilon \frac{m_{e\tau}^2 - m^2_{e\mu}}{m^3} \\
\frac{m_{e\mu}}{\sqrt{2} m} + \epsilon\, 
  \frac{ m_{e\tau}(2m_{e\tau}^2 -m_{e\mu}^2 )}{2 \sqrt{2} m^4} &
- \frac{m_{e\mu}}{\sqrt{2} m} + \epsilon\, 
  \frac{ m_{e\tau}(2m_{e\tau}^2 -m_{e\mu}^2 )}{2 \sqrt{2} m^4} &
- \frac{m_{e\tau}}{m} \\
\frac{m_{e\tau}}{\sqrt{2} m} + \epsilon\, 
  \frac{ m_{e\mu}(2m_{e\mu}^2 -m_{e\tau}^2 )}{2 \sqrt{2} m^4} &
- \frac{m_{e\tau}}{\sqrt{2} m} + \epsilon\, 
  \frac{ m_{e\mu}(2m_{e\mu}^2 -m_{e\tau}^2 )}{2 \sqrt{2} m^4} &
 \frac{m_{e\mu}}{m} 
  \end{array}
 \right ) \nonumber 
\end{eqnarray}
where $m=\sqrt{m_{e\mu}^2 + m^2_{e\tau}}$.  The above results reduce to
the results of Ref. \cite{otto} in the limit $\epsilon\to 0$.

The mass eigenvalues $m_1,m_2,m_3$ and $U$ are arranged in the inverted
mass hierarchy such that 
$m_1^2 \simeq m_2^2 >> m_3^2$.  The mass-square difference between $m_1^2$ and
$m_2^2$ explains the solar neutrino while that between $m_2^2$ and
$m_3^2$ explains the atmospheric neutrino.  The mixing angles can also be
obtained from the matrix $U$ such that $\tan\theta_{\rm sol} \equiv
\tan\theta_{12} = U_{e2}/U_{e1}$ and $\tan\theta_{\rm atm} \equiv
\tan\theta_{23} = U_{\mu 3}/U_{\tau 3}$.
According to the most recent fit to the atmospheric neutrino data 
\cite{atm-fits} in Eq. (\ref{atm}),  the $\sin^2 2\theta_{\rm atm}$
is practically 1 and we obtain
\begin{equation}
\frac{m_{e\tau}}{m_{e\mu}} = \pm 1 \;\;\; {\rm and} \;\;\;
|m_{e\mu}| = |m_{e\tau}|= \sqrt{ \Delta m^2_{\rm atm}/2} = 0.039 \;{\rm eV} \;.
\end{equation}
The condition $|m_{e\mu}| = |m_{e\tau}|$ also implies $U_{e3} =0$, 
which is consistent with the CHOOZ limit \cite{chooz}.
The parameter $\epsilon$ is then given by
\begin{equation}
\epsilon = \frac{\Delta m^2_{\rm sol}}{2 \sqrt{2} m_{e\mu}} = \left \{
\begin{array}{ll}
   4.1 \times 10^{-4} \;{\rm eV} & \mbox{for LMA} \\
   4.2 \times 10^{-9} \;{\rm eV} & \mbox{for VAC} 
 \end{array} \right. \;.
\end{equation}
We note that the mixing angle $\tan\theta_{12} = 1+ \epsilon/(2^{3/2} m_{e\mu})
$ essentially equals 1, which is obviously incompatible with the SMA solution,
but fairly consistent with the LMA and VAC solutions \cite{solar-fits}.
Since the LMA solution is the more favorable one, at least statistically, and
moreover the required $\epsilon$ for the VAC solution is a much more fine-tuned
one, we shall concentrate our discussion on the LMA solution 
in the next section.

We can use Eq. (\ref{mab}) to obtain the values on $f_{ab}$ from the fitted
values of $m_{e\mu},m_{e\tau},m_{\mu\tau}$.  The ratio 
$m_{e\mu}:m_{e\tau}:m_{\mu\tau}=1:1:\Delta m^2_{\rm sol}/(\sqrt{2} \Delta 
m^2_{\rm atm})$, which implies
\begin{eqnarray}
f_{e\mu}\,:\,f_{e\tau}\,:\,f_{\mu\tau}&=& 1\,:\, \frac{m^2_\mu}{m^2_\tau}\,:\, 
\frac{m^2_\mu}{m^2_\tau} \, \frac{\Delta m^2_{\rm sol}}{
\sqrt{2} \Delta m^2_{\rm atm}}
\nonumber \\
&=& 1\,:\, 3.5\times 10^{-3} \,: \, 3.7\times 10^{-5} \;\;\; (3.8\times10^{-10}
) \;,
\end{eqnarray}
where the numerical value in the parenthesis is for the VAC solution.
Taking $M_{h_d^-},M_{h^-_{\rm Zee}}, m_3 \sim 100$ GeV, we obtain 
$f_{e\mu} \sim 10^{-5} - 10^{-4}$, $f_{e\tau} \sim 10^{-7}$, and 
$f_{\mu\tau} \sim 10^{-9}$ for the LMA solution.

These $f_{ab}$'s are also subject to other constraints coming muon and
tau decays \cite{mituda}.  
The decay $\mu \to e \bar \nu \nu$ puts a bound on
$f_{e\mu}$: $|f_{e\mu}|^2/M^2 < 7\times 10^{-4} G_F$, where $M$ is a common
mass scale for $M_{h_d^-},M_{h^-_{\rm Zee}}$ and $G_F$ is the Fermi constant.
Taking $M\sim 100$ GeV, we obtain $|f_{e\mu}| < 9\times 10^{-3}$, which is
consistent with the $f_{e\mu}$ obtained above to explain the neutrino mass.
The branching ratio $B(\mu\to e\gamma)$ puts a bound on
$|f_{\mu\tau} f_{e\tau}|/M^2 < 2.8\times 10^{-4} G_F$, which implies
$|f_{\mu\tau} f_{e\tau}| <3 \times 10^{-5}$ for $M\sim 100$ GeV.
This is also consistent with the $f_{\mu\tau}, f_{e\tau}$ obtained above.

\section{Patterns predicted by $0\nu\beta\beta$ data}

If the $0\nu\beta\beta$ signal is true, it demands a very large value for
$m_{ee}$ relative to the mass-square differences required for the atmospheric
and solar neutrino data.  It has been shown \cite{klap2,xing,mina,hambye} 
that 
the most favorable mass pattern is a degenerate one with small mass differences
among the mass eigenstates.
\footnote{To explain the $0\nu\beta\beta$ data alone it is not necessary to
invoke neutrino oscillations.  For example, $R$-parity violation \cite{uehara}
and other new physics such as leptoquarks \cite{klap3} can explain 
the $0\nu\beta\beta$ data.} 

To obtain the mass pattern it is easy to work bottom-up.  We work in the basis
where the charged-lepton mass matrix is diagonal such that the mixing 
information is entirely contained in the neutrino mass matrix. From the data
we can write down specific forms of the mixing matrix $U$ and the diagonal
mass matrix $M_D$ in the mass basis, then we can obtain the
mass matrix $M_\nu$ in the flavor basis as
\begin{equation}
\label{mnu}
M_\nu = U\; M_D \;U^T \;,
\end{equation}
where the flavor basis and mass basis are related by
\begin{equation}
\left(  \begin{array}{c}
            \nu_e \\
            \nu_\mu \\
            \nu_\tau \end{array} \right ) 
  =
\left(  \begin{array}{ccc}
            U_{e1} & U_{e2} & U_{e3} \\
            U_{\mu 1} & U_{\mu 2} & U_{\mu 3} \\
            U_{\tau 1} & U_{\tau 2} & U_{\tau 3} \end{array} \right ) \;\;
\left(  \begin{array}{c}
            \nu_1 \\
            \nu_2 \\
            \nu_3 \end{array} \right ) \;.
\end{equation}
Assuming no CP violation a general form of $U$ is given by
\begin{equation}
U = U_{23} \; U_{13} \; U_{12}\;,
\end{equation}
where $U_{ij}$ is the rotation matrix about the $i$ and $j$'th mass 
eigenstates.  By considering the various data sets the general form of 
$U$ can be reduced to a simple form.
Since we know that the
atmospheric neutrino requires a maximal mixing between the $\nu_\mu$ and
$\nu_\tau$, $U_{23}$ is given by
\[
U_{23}= \left(  \begin{array}{ccc}
          1               &     0           & 0  \\
          0 &   \frac{1}{\sqrt 2} & -\frac{1}{\sqrt 2}  \\
          0 &   \frac{1}{\sqrt 2} &  \frac{1}{\sqrt 2}
            \end{array} \right ) \;.
\]
Since the angle $\theta_{13}$ is constrained by CHOOZ \cite{chooz} to be small,
we simply set $U_{13}=I$.  For the rotation between the $(1,2)$ states 
there are a few possible solutions to the solar neutrino, and so we used a 
generic $U_{12}$ as
\[
U_{12} = \left(  \begin{array}{ccc}
          c              &     s           & 0  \\
          -s &   c  & 0\\
          0 &    0 & 1
            \end{array} \right ) \;,
\]
where $c=\cos\theta_{12}, s=\sin\theta_{12}$. 
Therefore, $U$ is given by
\begin{equation}
\label{U}
U= \left(  \begin{array}{ccc}
          c               &     s           & 0  \\
          - \frac{s}{\sqrt 2} &   \frac{c}{\sqrt 2} & -\frac{1}{\sqrt 2}  \\
          - \frac{s}{\sqrt 2} &   \frac{c}{\sqrt 2} &  \frac{1}{\sqrt 2}
            \end{array} \right ) \;.
\end{equation}

For the mass matrix in the mass basis there are two cases: (i) normal and
(ii) inverted mass hierarchies, which we shall consider in turns.  In the 
 normal hierarchy, the solar oscillation is between the two lighter states,
 while the atmospheric oscillation is between the heaviest and the lighter
 states.  Recall the convention that the solar oscillation is between the ``1''
 and ``2'' states, we put the diagonal mass matrix as 
\begin{equation}
 \label{md1}
 M_D^{\rm normal} = \left(  \begin{array}{ccc}
           m_0               &    0           & 0  \\
           0   &   (m_0 + \delta) \, e^{i \phi}  & 0  \\
           0   &   0             &  (m_0 + m) \, e^{i \phi'}
             \end{array} \right ) \;,
 \end{equation}
 where $\delta \simeq \Delta m^2_{\rm sol}/(2 m_0)$ ,
       $m \simeq \Delta m^2_{\rm atm}/(2 m_0)$, and for $\phi$ and $\phi'$
 we take as 0 for the moment.  We shall later show the results
 for other values of $\phi$ and $\phi'$.
 Using Eqs. (\ref{mnu}), (\ref{U}), and (\ref{md1}), we obtain the neutrino
 mass matrix in the flavor basis:
\begin{equation}
M_\nu^{\rm normal} = 
\left(  \begin{array}{lll}
  m_0 + \delta\, s^2 & \frac{c s}{\sqrt{2}}\, \delta & 
                       \frac{c s}{\sqrt{2}}\, \delta \\
  \frac{c s}{\sqrt{2}}\, \delta   & m_0 + \frac{m}{2} + \frac{c^2}{2}\,\delta &
   -\frac{m}{2} + \frac{c^2}{2}\,\delta  \\
  \frac{c s}{\sqrt{2}}\, \delta &  -\frac{m}{2} + \frac{c^2}{2}\,\delta &
   m_0 + \frac{m}{2} + \frac{c^2}{2}\,\delta  
        \end{array}  
 \right ) \;.
\end{equation}
With input values  of $m_0=0.4$ eV, $\delta =5.6\times 10^{-5}$ eV, 
$m=3.75\times 10^{-3}$ eV, $c=1/\sqrt{1.41}, s=\sqrt{0.41/1.41}$, we obtain
\begin{eqnarray}
\Delta m^2_{32} \approx \Delta m_{31}^2 = 3\times 10^{-3}\;{\rm eV}^2\,, &&
\quad \sin^2 2\theta_{23} =1 \nonumber  \\
\Delta m^2_{21} \approx 4.5\times 10^{-5}\;{\rm eV}^2\,, &&
\quad \tan^2 \theta_{12} = 0.41 \nonumber  
\end{eqnarray}
which are exactly the best fits shown in Eqs. (\ref{atm}) and (\ref{solar}).
This is obvious because the input values for the parameters that we chose are
simply based on the data.  In other words, with the data we can fit the 
parameters of the model.

For the case of inverted mass hierarchy the solar neutrino oscillation is
between the two heavier states while the atmospheric neutrino oscillation
is between the lightest and the heavier states.  Recall the 
convention again that the solar oscillation is always between the ``1'' and
``2'' states, we used a $M_D^{\rm inverted}$:
\begin{equation}
 \label{md2}
 M_D^{\rm inverted} = \left(  \begin{array}{ccc}
           m_0 + m + \delta               &    0           & 0  \\
           0   &   m_0 + m  & 0  \\
           0   &   0        &  m_0 
             \end{array} \right ) \;,
 \end{equation}
in which we have chosen the phase angles to be zero.  Using Eqs. 
(\ref{mnu}), (\ref{U}), and (\ref{md2}) we obtain 
\begin{equation}
M_\nu^{\rm inverted} = 
\left(  \begin{array}{lll}
  m_0 + m+\delta\, c^2 & - \frac{c s}{\sqrt{2}}\, \delta & 
                         - \frac{c s}{\sqrt{2}}\, \delta \\
 -\frac{c s}{\sqrt{2}}\, \delta   & m_0 + \frac{m}{2} + \frac{s^2}{2}\,\delta &
   \frac{m}{2} + \frac{s^2}{2}\,\delta  \\
 -\frac{c s}{\sqrt{2}}\, \delta &  \frac{m}{2} + \frac{s^2}{2}\,\delta &
   m_0 + \frac{m}{2} + \frac{s^2}{2}\,\delta  
        \end{array}  
 \right ) \;.
\end{equation}
With the same input values for $m_0,\delta, m, c, s$ we obtain
\begin{eqnarray}
\Delta m^2_{32} \approx \Delta m_{31}^2 = - 3\times 10^{-3}\;{\rm eV}^2\,, &&
\quad \sin^2 2\theta_{23} =1 \nonumber  \\
\Delta m^2_{21} \approx - 4.5\times 10^{-5}\;{\rm eV}^2\,, &&
\quad \tan^2 \theta_{12} = 0.41 \nonumber  
\end{eqnarray}

In general, we can use the data to fit the values of $m_0, m, \delta, 
\theta_{12}$:
\begin{eqnarray}
m_0 = m_{ee}\;; \quad m \simeq \Delta m^2_{\rm atm}/(2 m_0)\;; \nonumber \\
\delta \simeq \Delta m^2_{\rm sol}/(2 m_0)\;, \;
\theta_{12} = \theta_{\rm sol} \;.
\end{eqnarray}
We have already used a particular form of $U$ such that a maximal mixing is
always between the ``2'' and ``3'' states.

Next we are going to show  the results when we allow nonzero phases 
for $\phi$ and $\phi'$. For illustrations we use (ii) $\phi=\pi, \phi'=0$,
(iii) $\phi=0, \phi'=\pi$, and (iv) $\phi=\pi, \phi'=\pi$.  The case (i)
$\phi=\phi'=0$ has already been shown above.
The neutrino mass matrices in the flavor basis for normal and inverted
mass hierarchies are given, respectively, by
\widetext
\begin{eqnarray}
{\rm (ii)}\; &&\phi=\pi, \phi'=0:  \nonumber \\
&&\left(  \begin{array}{ccc}
 \cos2\theta_{12}m_0 - \delta\, s^2 & 
- \frac{\sin2\theta_{12}}{\sqrt{2}}m_0 - \frac{c s}{\sqrt{2}}\, \delta & 
- \frac{\sin2\theta_{12}}{\sqrt{2}}m_0 - \frac{c s}{\sqrt{2}}\, \delta \\
- \frac{\sin2\theta_{12}}{\sqrt{2}}m_0 - \frac{c s}{\sqrt{2}}\, \delta & 
-\frac{1}{2} \cos2\theta_{12}m_0 + \frac{m_0+m}{2} - \frac{c^2}{2}\delta & 
-\frac{1}{2} \cos2\theta_{12}m_0 - \frac{m_0+m}{2} - \frac{c^2}{2}\delta \\
- \frac{\sin2\theta_{12}}{\sqrt{2}}m_0 - \frac{c s}{\sqrt{2}}\, \delta & 
-\frac{1}{2} \cos2\theta_{12}m_0 - \frac{m_0+m}{2} - \frac{c^2}{2}\delta &
-\frac{1}{2} \cos2\theta_{12}m_0 + \frac{m_0+m}{2} - \frac{c^2}{2}\delta 
        \end{array}  
 \right )   \nonumber \\
&&
\left(  \begin{array}{ccc}
 \cos2\theta_{12}(m_0+m) +\delta\, c^2 & 
- \frac{\sin2\theta_{12}}{\sqrt{2}}(m_0+m) - \frac{c s}{\sqrt{2}}\, \delta & 
- \frac{\sin2\theta_{12}}{\sqrt{2}}(m_0+m) - \frac{c s}{\sqrt{2}}\, \delta \\
- \frac{\sin2\theta_{12}}{\sqrt{2}}(m_0+m) - \frac{c s}{\sqrt{2}}\, \delta & 
-\frac{1}{2} \cos2\theta_{12}(m_0+m) + \frac{m_0}{2} + \frac{s^2}{2}\delta & 
-\frac{1}{2} \cos2\theta_{12}(m_0+m) - \frac{m_0}{2} + \frac{s^2}{2}\delta \\
- \frac{\sin2\theta_{12}}{\sqrt{2}}(m_0+m) - \frac{c s}{\sqrt{2}}\, \delta & 
-\frac{1}{2} \cos2\theta_{12}(m_0+m) - \frac{m_0}{2} + \frac{s^2}{2}\delta &
-\frac{1}{2} \cos2\theta_{12}(m_0+m) + \frac{m_0}{2} + \frac{s^2}{2}\delta 
        \end{array}  
 \right )  \nonumber\\
{\rm (iii)}\; &&\phi=0, \phi'=\pi:  \nonumber \\
&&\left(  \begin{array}{ccc}
 m_0 + \delta\, s^2 & \frac{c s}{\sqrt{2}}\, \delta & 
                      \frac{c s}{\sqrt{2}}\, \delta \\
\frac{c s}{\sqrt{2}}\, \delta &
- \frac{m}{2} + \frac{c^2}{2}\delta & 
m_0 + \frac{m}{2} + \frac{c^2}{2}\delta \\
\frac{c s}{\sqrt{2}}\, \delta &
m_0 + \frac{m}{2} + \frac{c^2}{2}\delta &
- \frac{m}{2} + \frac{c^2}{2}\delta  \\
        \end{array}  
 \right )   \nonumber \\
&&\left(  \begin{array}{ccc}
 m_0 + m + \delta\, c^2 & -\frac{c s}{\sqrt{2}}\, \delta & 
                          -\frac{c s}{\sqrt{2}}\, \delta \\
-\frac{c s}{\sqrt{2}}\, \delta &
 \frac{m}{2} + \frac{s^2}{2}\delta & 
m_0 + \frac{m}{2} + \frac{s^2}{2}\delta \\
-\frac{c s}{\sqrt{2}}\, \delta &
m_0 + \frac{m}{2} + \frac{s^2}{2}\delta &
 \frac{m}{2} + \frac{s^2}{2}\delta  \\
        \end{array}  
 \right )   \nonumber \\
{\rm (iv)}\; &&\phi=\pi, \phi'=\pi:  \nonumber \\
&&\left(  \begin{array}{ccc}
 \cos2\theta_{12}m_0 - \delta\, s^2 & 
- \frac{\sin2\theta_{12}}{\sqrt{2}}m_0 - \frac{c s}{\sqrt{2}}\, \delta & 
- \frac{\sin2\theta_{12}}{\sqrt{2}}m_0 - \frac{c s}{\sqrt{2}}\, \delta \\
- \frac{\sin2\theta_{12}}{\sqrt{2}}m_0 - \frac{c s}{\sqrt{2}}\, \delta & 
-\frac{1}{2} \cos2\theta_{12}m_0 - \frac{m_0+m}{2} - \frac{c^2}{2}\delta & 
-\frac{1}{2} \cos2\theta_{12}m_0 + \frac{m_0+m}{2} - \frac{c^2}{2}\delta \\
- \frac{\sin2\theta_{12}}{\sqrt{2}}m_0 - \frac{c s}{\sqrt{2}}\, \delta & 
-\frac{1}{2} \cos2\theta_{12}m_0 + \frac{m_0+m}{2} - \frac{c^2}{2}\delta &
-\frac{1}{2} \cos2\theta_{12}m_0 - \frac{m_0+m}{2} - \frac{c^2}{2}\delta 
        \end{array}  
 \right )   \nonumber \\
&&
\left(  \begin{array}{ccc}
 \cos2\theta_{12}(m_0+m) +\delta\, c^2 & 
- \frac{\sin2\theta_{12}}{\sqrt{2}}(m_0+m) - \frac{c s}{\sqrt{2}}\, \delta & 
- \frac{\sin2\theta_{12}}{\sqrt{2}}(m_0+m) - \frac{c s}{\sqrt{2}}\, \delta \\
- \frac{\sin2\theta_{12}}{\sqrt{2}}(m_0+m) - \frac{c s}{\sqrt{2}}\, \delta & 
-\frac{1}{2} \cos2\theta_{12}(m_0+m) - \frac{m_0}{2} + \frac{s^2}{2}\delta & 
-\frac{1}{2} \cos2\theta_{12}(m_0+m) + \frac{m_0}{2} + \frac{s^2}{2}\delta \\
- \frac{\sin2\theta_{12}}{\sqrt{2}}(m_0+m) - \frac{c s}{\sqrt{2}}\, \delta & 
-\frac{1}{2} \cos2\theta_{12}(m_0+m) + \frac{m_0}{2} + \frac{s^2}{2}\delta &
-\frac{1}{2} \cos2\theta_{12}(m_0+m) - \frac{m_0}{2} + \frac{s^2}{2}\delta 
        \end{array}  
 \right )  \nonumber
\end{eqnarray}

\endwidetext

The present data cannot distinguish the above patterns because they all
give correct $\Delta m^2_{\rm atm}$, $\sin^2 2\theta_{\rm atm}$,
$\Delta m^2_{\rm sol}$, $\tan^2 \theta_{\rm sol}$, and $m_{ee}$ by adjusting
the parameters $m_0,m,\delta, \theta_{12}$.  The preference of one pattern 
over the others is a matter of taste.  We suggest two fine-tuning criteria:
(1) the smaller the ratios $m/m_0$ and $\delta/m_0$ the worse the 
fine-tuning is,
and (2) the stability of the maximal mixing provided by the mass matrix.
The first criterion prefers cases (i) and (iii) because cases (ii) and (iv)
require a larger $m_0$ in order to explain the $0\nu\beta\beta$ data.
Second, the mass matrices in (i) are not stable, which is explained as follows.
Let us look at the $(2,3)$ block of the matrix, which is of the form
$\left(\begin{array}{cc}
          1 & \epsilon \\
        \epsilon & 1 \end{array} \right )$, where $\epsilon \ll 1$.  The
maximal mixing between the ``2'' and ``3'' states is very unstable.  Once
$m_{\mu\mu}$ and $m_{\tau\tau}$ differ very slightly, the maximal mixing
is destroyed.  We have verified that if $m_{\mu\mu}$ or $m_{\tau\tau}$ 
increases by 1\% the resulting $\sin^2 2\theta_{\rm atm} \sim 0.5$, already 
out of the allowed range.  It is barely consistent with the data range if
$m_{\mu\mu}$ or $m_{\tau\tau}$ is changed by less than 0.5\%.  
On the other hand, in case (iii) the $(2,3)$ block is of
the form
$\left(\begin{array}{cc}
           \epsilon & 1 \\
      1&  \epsilon  \end{array} \right )$.
This is a rather robust form such that even when $m_{\mu\mu}$ 
or $m_{\tau\tau}$ is changed appreciably, the maximal mixing remains.  
We have checked even if $m_{\mu\mu}$ or $m_{\tau\tau}$ increases by a factor
of two the maximal mixing solution remains for the atmospheric neutrino.
This is easy to understand.  As long as $m_{\mu\tau} \gg 
m_{\mu\mu}, m_{\tau\tau}$ we have the maximal mixing in the $(2,3)$ block.
So we conclude this section by saying that the mass matrices in case (iii) are
the more favorable ones in view of stability and fine tuning.

Since in order to explain the $0\nu\beta\beta$ data a nonzero $m_{ee}$ is 
needed, which immediately rules out the original Zee model.  In the next 
section, we look at a few extensions that can explain
the $0\nu\beta\beta$ data while maintaining the solutions to atmospheric
and solar neutrino problems.  

We end this section by saying that 
we have checked the patterns given in Ref. \cite{klap2}.  We found that we
can reproduce their $\Delta m^2_{\rm atm}$, $\Delta m^2_{\rm sol}$, and
$\sin^2 2\theta_{\rm atm}$, but not $\sin^2 2\theta_{\rm sol}$.  We believe
they calculated the solar mixing angle using the wrong pair of states, instead
of using the pair whose mass-square difference is $\Delta m^2_{\rm sol}$.

\section{Extensions to the Zee model}

Since the contributions to neutrino mass matrix in the Zee model arise from 
1-loop diagrams, it is natural to consider the 2-loop contributions as well.
It was pointed out \cite{darwin} that some 2-loop generalizations of 
the Zee model may be able to generate nonzero diagonal matrix elements.
However, it is very difficult to accommodate the LMA solution unless by 
fine tuning of the model parameters \cite{arcadi}.  In the following, we
consider a couple of extensions that can hopefully accommodate the data in
a less unnaturally way, though fine tuning may still be necessary.

A straight-forward extension is to add a heavy right-handed neutrino $N_i$
for each generation with an interaction: ${\cal L}= -y_i L_i H_u N_i - 
\frac{M_i}{2} N_i  N_i +h.c.$, where $M_i$ is a mass scale much larger than 
the weak scale and $y_i$ is a Yukawa 
coupling.  After the electroweak symmetry breaking, the interaction gives
a Dirac mass term to the corresponding SM neutrino.  Through
the see-saw mechanism a small neutrino mass is then generated. 
Integrating out the heavy right-handed neutrinos, the effective
interaction term that gives rise to neutrino mass is given by
\begin{equation}
{\cal L} = y_i^2 \frac{ (L_i H_u) (L_i H_u)}{2 M_i} +h.c. \;,
\end{equation}
where we shall simply take a common scale for all $M_i=M$.  In this way
diagonal majorana mass terms can be generated easily as
\begin{equation}
m_{ii} = - y_i^2 \frac{ \langle H_u \rangle^2}{M}\;,
\end{equation}
for $i=e,\mu,\tau$.  
The appropriate values for the diagonal matrix elements displayed in the 
last section can be obtained by adjusting the $y_i$'s, though in some sense it 
may require a fine-tuning.  The off-diagonal matrix elements can be adjusted
by varying the parameters of the Zee model.

Another interesting extension to the Zee model is in fact a supersymmetrization
of the Zee model within the minimal supersymmetric model (MSSM) 
without $R$-parity conservation \cite{otto}.  The right-handed slepton works 
as the charged gauge singlet (Zee singlet) 
while the MSSM already has two Higgs 
doublets, which provide the mixing between the $H_d$ and Zee singlet through
the VEV of the $H_u$.  Here the $R$-parity violating terms $\mu_i L_i H_u$
provide mixings among neutrinos and higgsinos.   Through a see-saw type
mechanism tree-level contributions to neutrino masses arise and the
$m_{ij}$ are roughly proportional to $\mu_i \mu_j/\mu$, where $\mu$ 
is the Higgs parameter in the $\mu H_u H_d$ term.  Thus, the diagonal
matrix elements are nonzero.  There are also one-loop type contributions
to the matrix elements in both diagonal and off-diagonal elements. It is
then possible to adjust the soft and $R$-parity violating parameters in order
to obtain the mass matrix in (iii) of the last section.  Again, to some
extent fine-tuning of the parameters is needed to ensure the form of the
mass matrix.

\section{Conclusions}

We have examined the Zee model in light of the most recent fits to the
solar and atmospheric neutrino data.  The Zee model is a natural candidate
that can provide a bi-maximal mixing solution to the atmospheric and solar
neutrino problems.  The resulting atmospheric mixing angle $\theta_{23}$ 
is right at the best value of the experimental data while the solar mixing
angle $\theta_{12}$ is fairly consistent with the allowed range of 
the LMA and VAC solutions.  By fitting to the data we found that
the lepton-number-violating
coupling $f_{e\mu}$ is of order $10^{-5} - 10^{-4}$ and the ratio 
$f_{e\mu}\,:\,f_{e\tau}\,:\,f_{\mu\tau}=
1\,:\, 3.5\times 10^{-3} \,: \, 3.7\times 10^{-5}$ for the LMA solution.
The size of these couplings is consistent with other constraints coming 
from muon and tau decays.

The recent neutrinoless double beta decay \cite{0nbeta,vissani,aal}, 
if confirmed, rules out the
original Zee model, because the positive signal requires a nonzero $m_{ee}$
matrix element.  Moreover, a large $m_{ee}$ value necessarily gives large 
$m_{1,2,3}$ for the mass eigenstates.  Such neutrino masses form a 
significant fraction of the warm dark matter \cite{vernon}.
We have derived a few general patterns of neutrino mass matrix from the
observed data of neutrino oscillations (here we assumed the LMA solution
for the solar 
neutrino.)  Due to fine-tuning and stability reasons the forms in (iii) of
Sec. III are preferred.  
Nevertheless, the interpretation of the $0\nu\beta\beta$ signal requires 
a consensus among various experimental groups before generally accepted 
by the community. 

Finally, we offered a couple of extensions to the Zee model that can give
nonzero diagonal matrix elements.  However, in order to obtain the 
appropriate mass patterns fine-tuning of the parameters involved 
is still necessary.

We are grateful to Otto Kong for useful discussions. 
This research was supported in part by the National Center for Theoretical
Science under a grant from the National Science Council of Taiwan R.O.C.



\begin{thebibliography}{99}

\bibitem{superk}
Super-Kamiokande Collaboration, S. Fukuda {\it et al.}, Phys. Rev. Lett.
{\bf 81}, 1562 (1998); Phys. Rev. Lett. {\bf 82}, 2644 (1999); 
Phys. Rev. Lett. {\bf 85}, 3999 (2000).

\bibitem{k2k}
K2K Collaboration, in Proceedings of Snowmass 2001, Snowmass, Colorado, 
30 Jun - 21 Jul 2001, e-Print Archive: hep-ex/0110034.

\bibitem{atm-fits}
N. Fornengo, M. Gonzalez-Garcia, and J. Valle, Nucl. Phys. {\bf B580}, 
58 (2000); 
G. Fogli, E. Lisi, A. Marrone, and D. Montanino, 
Nucl. Phys. Proc. Suppl. {\bf 91}, 167 (2000), e-Print Archive: hep-ph/0009269.

\bibitem{sno}
SNO Collaboration, Q.R. Ahmad {\it et al.}, Phys. Rev. Lett. {\bf 87}, 071301
(2001).

\bibitem{solar-fits}
G.L. Fogli, E. Lisi, D. Montanino, and A. Palazzo, Phys.Rev. {\bf D64}, 093007
(2001);
J. Bahcall, M. Gonzalez-Garcia, and C. Pena-Garay, JHEP {\bf 0108}, 014 (2001).

\bibitem{lsnd}
LSND Collaboration, A. Aguilar {\it et al.}, Phys. Rev. {\bf D64}, 112007
(2001); Phys. Rev. Lett. {\bf 81}, 1774 (1998).

\bibitem{0nbeta}
H.V. Klapdor-Kleingrothaus {\it et al.}, Mod. Phys. Lett. {\bf A16}, 2409 (2002).

\bibitem{vissani}
F. Feruglio, A. Strumia, and F. Vissani, e-Print Archive: hep-ph/020129.

\bibitem{aal}
C.E. Aalseth et al., hep-ex/0202018.

\bibitem{zee}
A. Zee, Phys. Lett. {\bf B93}, 389 (1980), Erratum-ibid. {\bf B95}, 461 (1980).

\bibitem{jarl}
C. Jarlskog, M. Matsuda, S. Skadhauge, M. Tanimoto, Phys. Lett. {\bf B449}, 
240 (1999).

\bibitem{paul1}
P. Frampton and S. Glashow, Phys. Lett. {\bf B461}, 95 (1999).

\bibitem{otto}
K. Cheung and O.C.W. Kong, Phys. Rev. {\bf D} (2000).

\bibitem{paul2}
P. Frampton, M. Oh, and T. Yoshikawa, e-Print Archive: hep-ph/0110300.

\bibitem{koide1}
Y. Koide, Phys. Rev. {\bf D64}, 077301 (2001).

\bibitem{haba}
N. Haba, K. Hamaguchi, and T. Suzuki, Phys. Lett. {\bf B519}, 243 (2001).

\bibitem{brah}
B. Brahmachari and S. Choubey, e-Print Archive: hep-ph/0111133.

\bibitem{dan1}
P. Frampton, S. Glashow, D. Marfatia, e-Print Archive: hep-ph/0201008.

\bibitem{petcov}
S. Petcov, Phys. Lett. {\bf B115}, 401 (1982).

\bibitem{john}
D. Dicus, H.-J. He, and J. Ng, Phys. Rev. Lett. {\bf 87}, 111803 (2001).

\bibitem{koide2}
Y. Koide and A. Ghosal, Phys. Rev. {\bf D63}, 037301 (2000).

\bibitem{chooz}
CHOOZ Collaboration, M. Apollonio {\it et al.}, Phys. Lett. {\bf B338}, 383
(1988); Phys. Lett. {\bf B466}, 415 (1999).

\bibitem{mituda}
A. Smirnov and M. Tanimoto, Phys. Rev. {\bf D55}, 1665 (1997);
A. Ghosal, Y. Koide, and H. Fusaoka, Phys. Rev. {\bf D64}, 053012 (2001);
E. Mituda and K. Sasaki, e-Print Archive: hep-ph/0103202.

\bibitem{klap2}
H.V. Klapdor-Kleingrothaus and U. Sarkar, e-Print Archive: hep-ph/0202006.

\bibitem{xing}
Z.-Z. Xing, e-Print Archive: hep-ph/0202034.

\bibitem{mina}
H. Minakata and H. Sugiyama, e-Print Archive: hep-ph/0202003 

\bibitem{hambye}
T. Hambye, e-Print Archive: hep-ph/0201307.

\bibitem{uehara}
Y. Uehara, e-Print Archive: hep-ph/0201277.

\bibitem{klap3}
H.V. Klapdor-Kleingrothaus and U. Sarkar, Mod. Phys. Lett. {\bf A16}, 2469
 (2001).

\bibitem{darwin}
D. Chang and A. Zee, Phys. Rev. {\bf D61}, 071303 (2000).

\bibitem{arcadi}
J. Oliver and A. Santamaria, Phys. Rev. {\bf D65}, 033003 (2002).

\bibitem{vernon}
V. Barger, S.L. Glashow, D. Marfatia, and K. Whisnant, 
e-Print Archive: hep-ph/0201262.

\end{thebibliography}
\end{document}